\documentclass[11pt]{article}
\usepackage{amsmath,amssymb,amsthm}
\usepackage{epsfig}
\usepackage{setspace}
\usepackage{a4wide}

\newtheorem{theorem}{Theorem}[section]
\newtheorem{proposition}{Proposition}[section]

\newcommand{\eat}[1]{}
\begin{document}

\eat{
\pagestyle{empty}
\noindent
{\bf Contact Author:}\\
Floris Geerts\\
Laboratory for Foundations of Computer Science\\
School of Informatics \\
James Clerk Maxwell Building, Room 2602 \\
The King's Buildings \\
Mayfield Road \\
Edinburgh EH9 3JZ\\ 
Scotland, UK \\
Telephone: +44 (0)131 650 5139 \\
Fax: +44 (0)131 667-7209 \\
{\tt fgeerts@inf.ed.ac.uk}
\newpage

{\bf keywords:}
 topological simplification, network graph algorithms,
online algorithms.
\newpage

\pagestyle{plain}
\doublespace
\begin{titlepage}}

\title{On-line topological simplification \\ of weighted graphs}
\author{Floris Geerts~$^{1,3}$ \and Peter Revesz$^2$~\thanks{Work done while
on a sabbatical leave from the University of
    Nebraska-Lincoln.  Work supported in
    part by USA NSF grants IRI-9625055 and IRI-9632871.}
\and Jan Van den Bussche$^{1}$}
\date{$^1$~Hasselt University/Transnational University Limburg\\
$^2$~University of Nebraska-Lincoln\\
$^3$~University of Edinburgh}

\maketitle
 \begin{abstract}
  We describe two efficient on-line  algorithms to simplify weighted graphs by
  eliminating degree-two vertices.  Our algorithms are on-line in that 
  they react to updates on the data,
  keeping the simplification up-to-date.  The supported updates are
  insertions of vertices and edges; hence, our algorithms are
  partially dynamic.  We provide both analytical and
  empirical evaluations of the efficiency of our approaches.
  Specifically, we prove an $O(\log n)$  upper bound on the amortized
  time complexity of our maintenance algorithms, with $n$ the number of
  insertions.
\end{abstract}
%\end{titlepage}
\section{Introduction}
Many GIS applications involve data in the form of a network, such
as road, railway, or river networks. It is  common to represent network
data in the form of so-called {\em polylines}. A polyline consists of a
sequence of consecutive straight-line segments of  variable length.
Polylines allow for the modeling of both straight lines and 
curved lines. 
A point on a polyline in which exactly two straight-line segments meet, is called
a {\em regular} point. Regular points are important for the modeling of
curved lines. Indeed, to represent accurately a curved line by a polyline, one
needs to use many regular points. Curved lines often occur in river networks, or
in road networks over hilly terrain.

We illustrate this in Figure~\ref{Fig:regular}
in which we show a part of the road network in the Ardennes (Belgium). In this hilly region,
many bended roads occur. As can be seen in the Figure,
there is an abundance of regular points --- which is often the case in real
network maps~\cite{Wor95}.

\begin{figure}
\includegraphics[angle=-90,width=\textwidth]{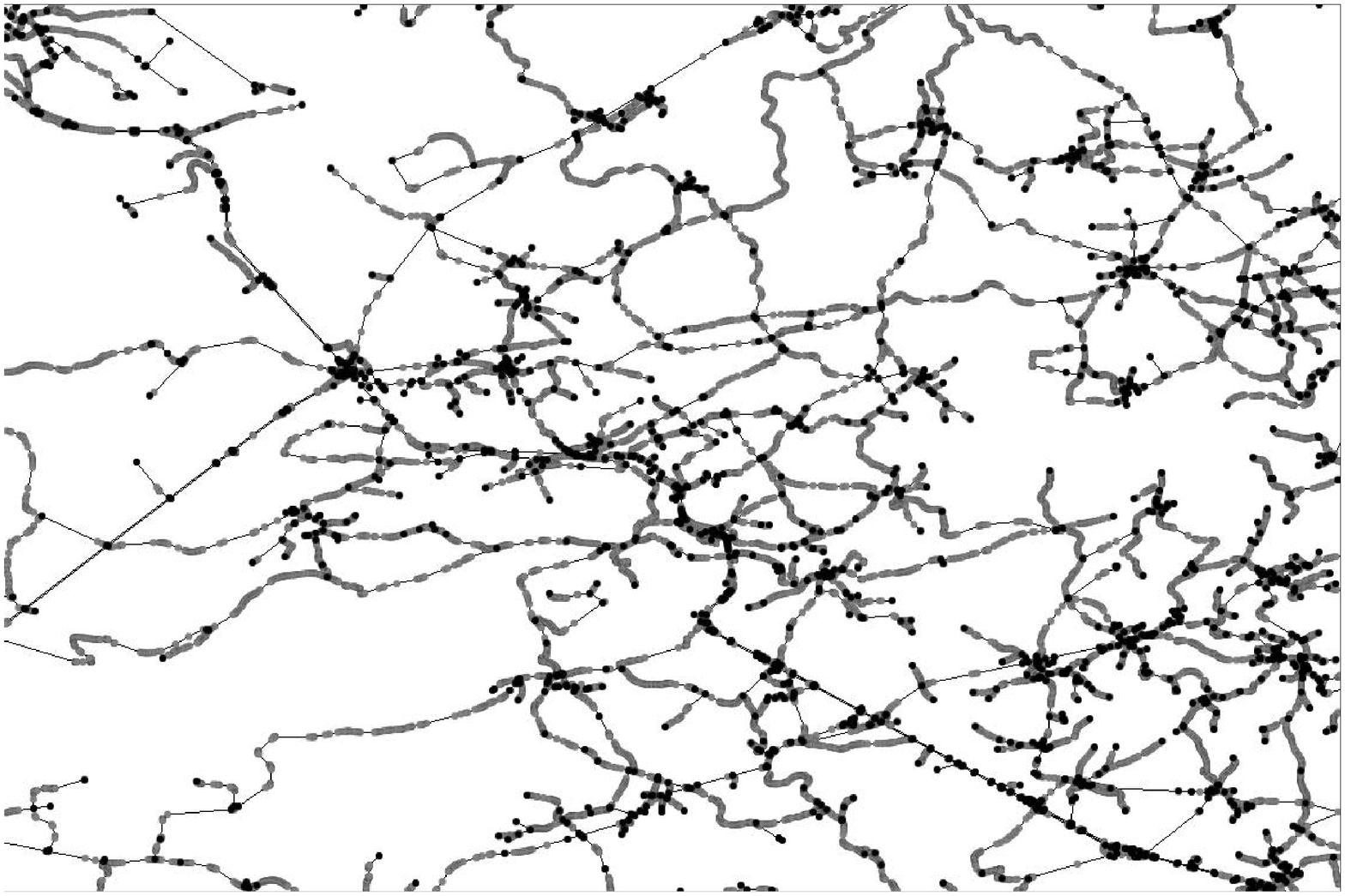}
\includegraphics[width=\textwidth]{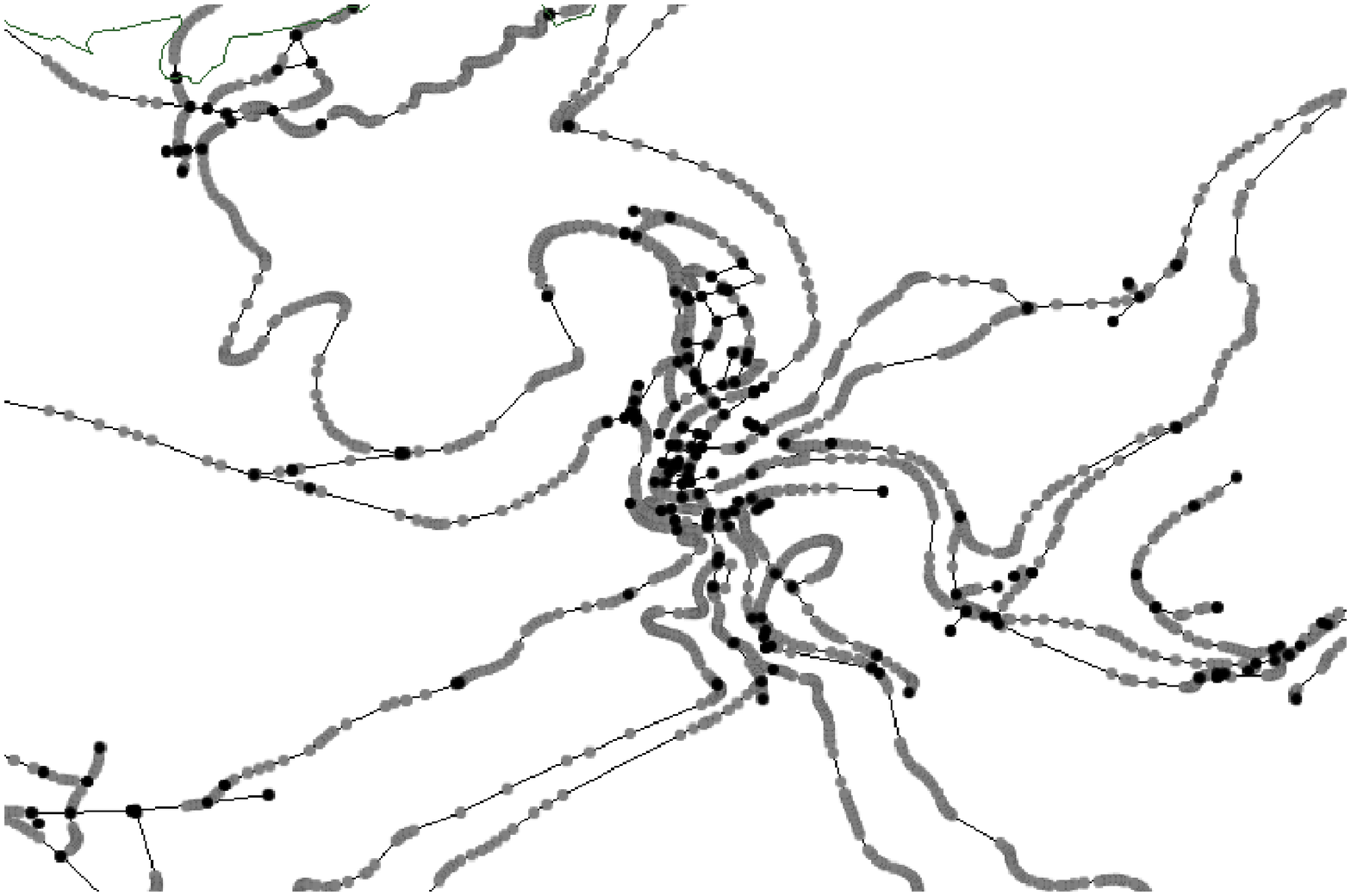}
\caption{Road network in two villages Tenneville (top) and La Roche (bottom) in the Belgian
Ardennes. Black points are non-regular points; gray points are regular. We see that the number of gray points is very high.}\label{Fig:regular}
\end{figure}

Although regular points are necessary to model the reality accurately, for
many applications they can be disregarded. More specifically, for
topological queries such as path queries,
one can ``topologically simplify'' the network by eliminating
all regular points; and answer the query (more efficiently) on the much
simplified network. Even when the network contains distance
information, one still can topologically simplify the network, but maintain
the distance information, as we will show in the present paper. More generally,
we work with arbitrary weight information.

\eat{
Topological queries such as
shortest paths or transitive closure, are very important in this
context. In this paper we will discuss a simplification method for
speeding up these queries on network-like data, represented as
undirected graphs with weights on the edges.

The simplification method consists of eliminating all ``regular''
vertices.  These are the vertices that simply lie on a line; in
graph-theoretic terms, they are the vertices of degree two. For
example, in Figure~\ref{Fig:regular}, a graph together with its
simplification are given.

Regular vertices occur often abundantly.  For example, in a road
network database, each bend in the road is represented by a
vertex, which will be regular.  The same applies more generally to
all networks represented on top of a discrete raster \cite{Wor95},
where a curved line is approximated by many straight line segments
between raster points, which then will be regular.

\begin{figure}
\centering \epsfig{height=4cm,file=giraf} \caption{A graph (thin
lines) together with its simplification (thick lines).}
\label{Fig:regular}
\end{figure}

}

Thus, the simplification of a network contains in a
compact manner the same topological and distance information as the original network. Such
``lossless topological representations'' have been studied by a number of
researchers \cite{kuijp,PSV,PODS}.  For example, initial
experiments reported by Segoufin and Vianu have shown drastic
compression of the size of the data by topological simplification. 
(The inclusion of distance information is new to the present paper.)

Of course, if we want to answer queries using the simplified
network instead of the original one, we are faced with the problem
of on-line maintenance of the simplified network under updates to
the original one. This problem is important due to
the dynamic character of certain network data. For example, suppose that there
 is a huge snowstorm which makes all roads unusable. As a result, 
 many snow clearing crews are sent to all parts of the city. They
continuously report back to a central station the road segments
that they have cleared.  The central station also continuously
updates its map of the usable network of roads. Moreover, big
arteries are cleared first, and therefore, the usable network will
have a high percentage of regular vertices in the initial stages.
While the snow is being cleared,  thousands of people may
query the database of the central station to find out what is the
shortest path they can take using the already cleared roads. 
Analogous applications requiring
on-line monitoring involve traffic jams in road networks, or 
downlinks in computer networks.

Two of us have reported on an initial investigation of this problem \cite{wij}.
The result was a maintenance algorithm that was {\em fully-dynamic},
 i.e., insertions and deletions of edges and vertices are allowed. This
 algorithm, however, is (in certain ``worst cases'') not any better
than redoing the simplification from scratch after every update, resulting
in an $O(n^2)$ time algorithm, where $n$ is the number of updates.
This is clearly not very practical.

The present paper proposes two very different algorithms for
on-line topological simplification:
\begin{enumerate}
\item {\bf Renumbering Algorithm}, which relies on the numbering and
renumbering of the regular vertices, takes on the average, only
{\em logarithmic\/} time per edge insertion to keep the simplified
network up-to-date; and
\item {\bf Topology Tree
Algorithm}, is based on the topology tree data structure of
Frederickson~\cite{fred1}  and has the same time complexity $O(n\log(n))$.
\end{enumerate}

Neither algorithm makes any assumptions on the graph, such as
planarity and the like. Real-life network data is often {\em
not\/} planar (e.g., in a road or railway network, bridges occur).
The presented algorithms are only {\em semi-dynamic}, in that they can
react efficiently to insertions (of vertices and edges), but not
to deletions. Insertions are sufficient for many applications (such as the
snow clearing mentioned above, were simply more and more road
segments become available again), but for applications requiring also
deletion,  the Topology Tree Algorithm easily can be extended to also
react correctly to edge deletions.

We have performed an empirical comparison of the Renumbering
Algorithm and the Topology Tree Algorithm using random, non-random
and two real data sets.

This paper is further organized as follows.  Basic definitions are
given in Section~2.  The general description of the on-line
simplification algorithm is described in Section~3. In Section~4,
we describe the Renumbering Algorithm, and in Section~5, the
Topology Tree Algorithm is described. The empirical comparison of
both algorithms is presented in Section~6.

\section{Basic Definitions}\label{sec:def}
Consider an undirected graph without self-loops
$G=(V,E,\lambda)$ with weighted edges;
the weights of the edges are given
by a mapping $\lambda:E\rightarrow {\bf R}^+$.
We will use the following definitions:
\begin{enumerate}
\item A vertex $v$ is {\em regular\/} if and only if it is adjacent to
precisely \index{regular vertex}  two edges.
\item A vertex that is not regular is called {\em
singular}.\index{singular  vertex}
\item A path between two singular vertices that passes only through
  regular vertices is called a {\em regular path}.
  \index{regular!path}
\end{enumerate}

We assume that the graph $G$ does not contain regular cycles:
cycles consisting of regular vertices only.

The {\em simplification\/} $G_s=(V_s,E_s,\lambda_s)$ of
$G$ is a multigraph with self-loops and weighted edges,
which is obtained as follows: (see Figure~\ref{Fig:regular})
\begin{enumerate}
\item $V_s$, the set of nodes of $G_s$, consists of all singular
vertices of $G$. \item $E_s$, the set of edges of $G_s$, formally
consists of all regular paths of $G$.  Every regular path between
two singular vertices $v$ and $w$ represents a \emph{topological
edge} in $G_s$ between $v$ and $w$. There might be multiple
regular paths between two singular vertices, so in general $G_s$
is a multigraph. \item the weight $\lambda_s(e)$ of a topological
edge $e$ is equal to the sum of all weights of edges on the
regular path corresponding to $e$.
\end{enumerate}

In the following, when a particular regular path $e$
between two singular vertices $v$ and
$w$ is clear from the context, we will abuse notation and conveniently
denote the topological edge $e$ by $\{v,w\}$.

\section{Online Simplification: General Description}\label{sec:update}

We consider only insertions of a new isolated vertex and
insertions of edges between existing vertices in the graph $G$
(other more complex insertion operations can be translated into a
sequence of these basic insertion operations).  The insertion of
an isolated vertex is handled trivially, i.e., we insert it into
$V_s$.

For the insertion of an edge we distinguish between
 six cases that are explained below.
The left side of each figure shows the
situation before the insertion of the edge $\{x,y\}$, drawn as the
dotted line, and the right side shows the situation after the
insertion.  The topological edges are drawn in thick
lines.
\begin{description}
\item[Case 1] Vertices $x$ and $y$ are both singular and $\deg(x)\neq 1$
  and $\deg(y)\neq 1$.
\begin{center}
\includegraphics[width=0.8\textwidth]{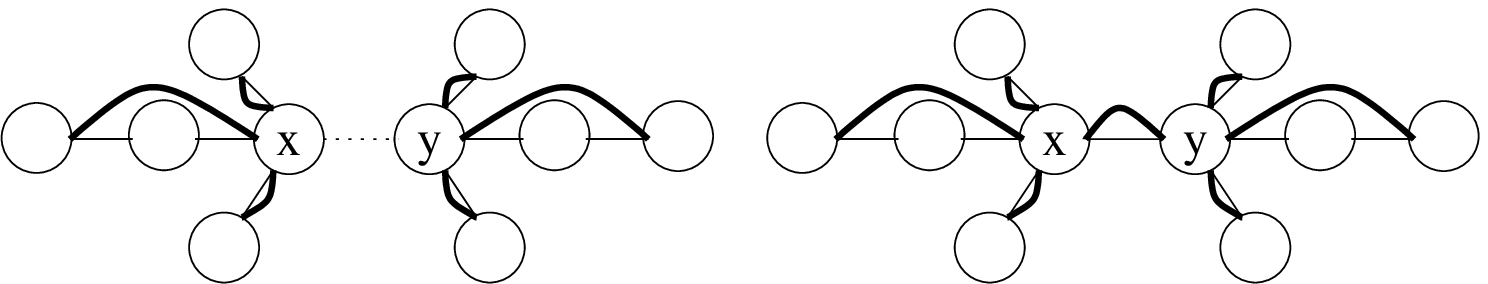}
\end{center}
Then the edge $\{x,y\}$ is also {\em inserted\/} in $G_s$.
\item[Case 2] Vertices $x$ and $y$ are both singular and one of them,
  say $x$, has degree one.
\begin{center}
\epsfig{width=0.8\textwidth,file=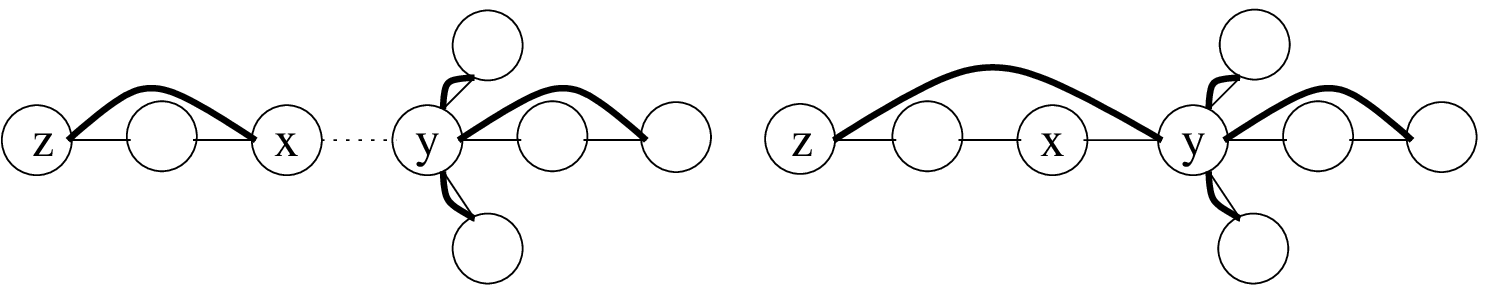}
\end{center}
Let $\{z,x\}$ be the edge in $G_s$ adjacent to $x$. {\em Extend\/}
  this edge  the new edge $\{z,y\}$ in $G_s$, putting
  $\lambda_s(\{z,y\}):=\lambda_s(\{z,x\})+\lambda(\{x,y\})$. Note that
  $x$ becomes a
  regular vertex after the insertion.
\item[Case 3]
  Vertices $x$ and $y$ are both singular and $\deg(x)=\deg(y)=1$.
\begin{center}
\epsfig{width=0.8\textwidth,file=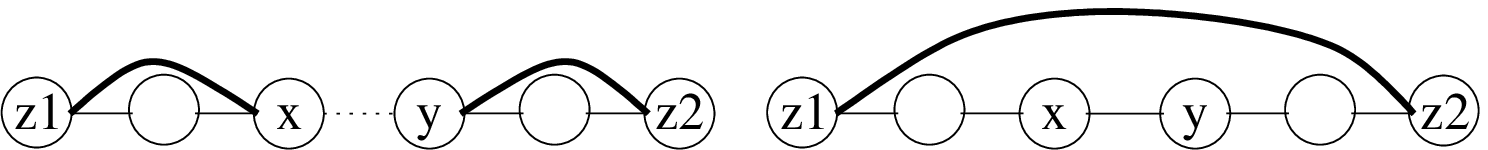}
\end{center}
  Let $\{z_1,x\}$ ($\{z_2,y\}$) be the edge in $G_s$ adjacent
  with $x$ ($y$). (Since we disallow regular cycles in $G$, we have
  $z_1\neq y$ and $z_2\neq x$.)
  Then {\em merge\/} the edges $\{z_1,x\}$ and $\{y,z_2\}$ in $G_s$ into
  a single, new edge $\{z_1,z_2\}$ in $G_s$, putting
  $\lambda_s(\{z_1,z_2\}):=\lambda_s(\{z_1,x\})+\lambda_s(\{y,z_2\})+\lambda(\{x,y\})$.
%  If $z_1=y$ ($z_2=x$) then {\em extend\/} the edge $\{x,y\}$
%  to, a new edge $\{x,x\}$ in $G_s$, putting
%  $\lambda_s(\{x,x\}):=\lambda_s(\{x,y\})+\lambda(\{x,y\})$.
\item[Case 4] One of the vertices $x$ and $y$ is regular, say $x$, and
  the other vertex, $y$, is singular and has degree one.
\begin{center}
\epsfig{width=0.8\textwidth,file=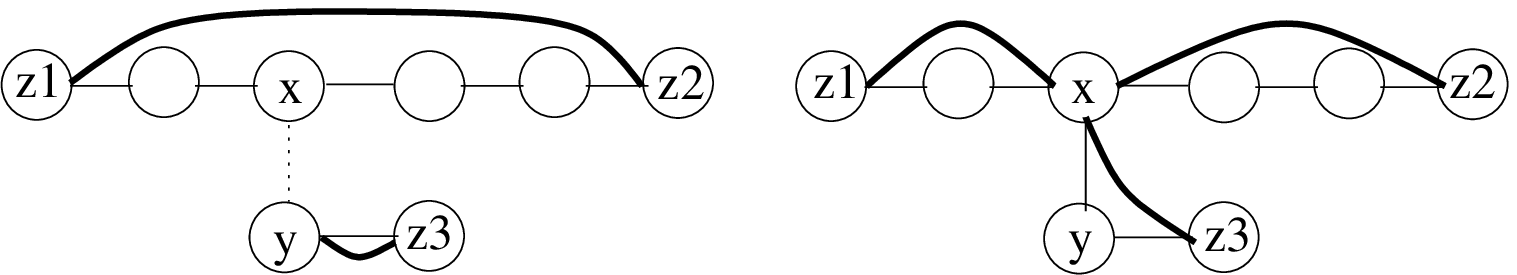}
\end{center}
 First, the edge $\{z_1,z_2\}$ of $G_s$ which
  corresponds to the regular path between $z_1$ and $z_2$ on which $x$
  lies, must be {\em split\/} into two
    new edges
  $\{z_1,x\}$ and $\{x,z_2\}$ of $G_s$. Here, we put
  $\lambda_s(\{z_1,x\}):=\sum \lambda(\{u,v\})$, where the summation is
  over all edges in $G$ on the regular path from $z_1$ to $x$. We
  similarly define $\lambda_s(\{x,z_2\})$. Secondly, let $\{z_3,y\}$ be
  the edge in $G_s$ adjacent to $y$. Then we {\em extend\/} this edge
  to a new edge $\{z_3,x\}$ in $G_s$, putting,
  $\lambda_s(\{x,z_3\}):=\lambda_s(\{y,z_3\})+\lambda(\{x,y\})$.

A special subcase occurs when $z_1=z_2$. In that case, the two paths from $x$
to $z_1$ give rise to two different edges from $x$ to $z_1$ in $G_s$ (recall
that $G_s$ was defined as a multigraph).
\item[Case 5] One of the vertices, say $x$, is regular and the other one,
  $y$, is singular with degree not equal to one.
\begin{center}
\epsfig{width=0.8\textwidth,file=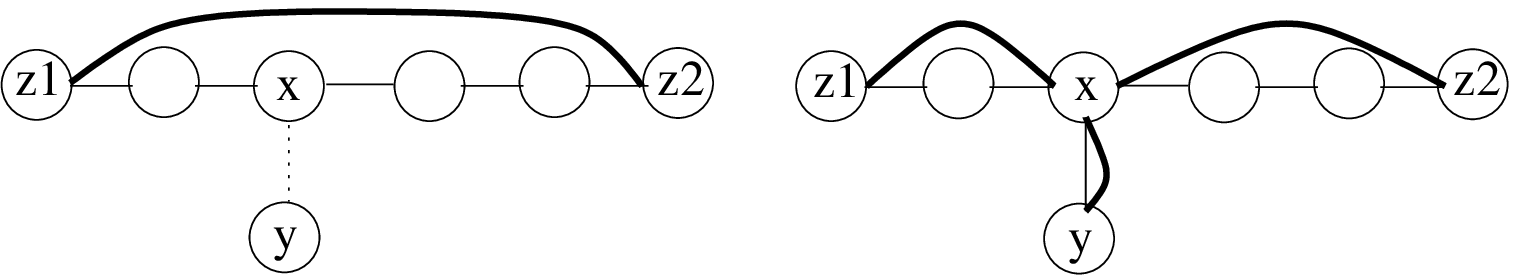}
\end{center}
  Then we split
  exactly as in case~4, and now we
  also insert $\{x,y\}$ as a new edge in $G_s$.
\item[Case 6] Both $x$ and $y$ are regular.
\begin{center}
\epsfig{width=0.8\textwidth,file=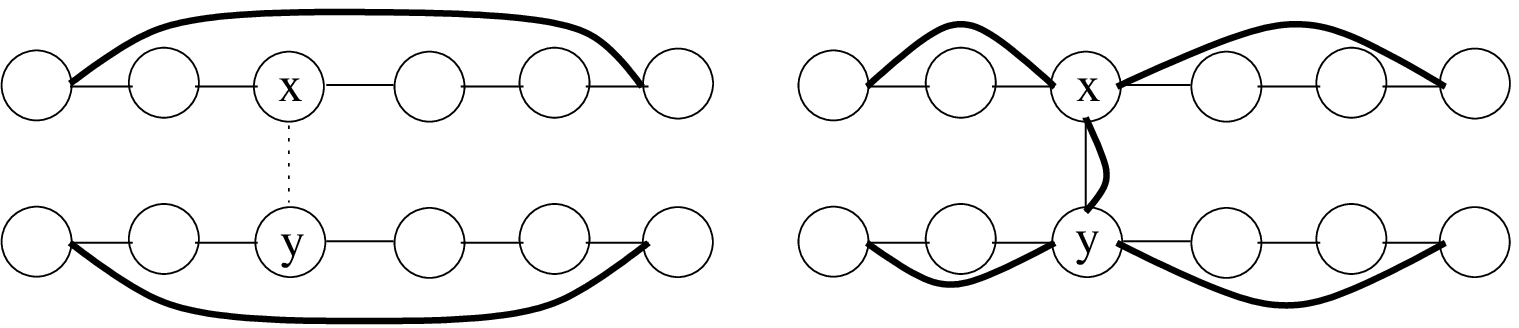}
\end{center}
Now, two {\em splits\/} must be performed.
\end{description}

As can be seen in the above description, if no regular vertices
are involved, then the update on the graph $G$ translates in a
straightforward way to an update on the simplification $G_s$. It
is only in cases 4, 5, and 6, that the update on the graph $G$
involves vertices which {\em have no counterpart\/} in the
simplification $G_s$. In these cases, we need to find the edge to
split and the weights of the topological edges created by the
split. Consequently, the problem of maintaining the simplification
$G_s$ of a graph $G$ amounts to two tasks:
\begin{itemize}
\item Maintain a function {\em find topological edge}, which takes a
 regular vertex as input, and outputs the topological edge whose corresponding
 regular   path in $G$ contains the input vertex.
\item Maintain a function {\em find weights\/} which outputs the weights
  of the edges created when a topological edge is split at the input
  vertex.
\end{itemize}

In an earlier, naive approach~\cite{wij},  we only discussed the
function {\em find topological edge.}
It worked by storing for each regular vertex a direct
pointer to its topological edge. This made the
topological edge accessible in constant time, but the maintenance
of the pointers under updates can be very inefficient in the worst
case.
 We next describe two algorithms
which are more efficient. Both algorithms keep the simplification
of a graph up-to-date
 when the graph is subject to edge insertions only.

\section{Online Simplification: Renumbering Algorithm}\label{sec:renum}
In this
section we introduce an algorithm for keeping the
simplification of a graph up-to-date when this graph is subject to
edge insertions. We first show how the topological edges can be found efficiently.

\subsection{Assigning numbers to the regular vertices} We number
\index{regular vertex!number}the regular vertices, that lie on a
regular path, consecutively. The numbers of the regular vertices
on any regular path will always form an interval of the natural
numbers. The Renumbering Algorithm  will maintain two properties:
\begin{description}
\item[Interval property:] the assignment of {\em consecutive\/} numbers
  to {\em consecutive\/} regular points;
\item[Disjointness property:] {\em different\/} regular paths have {\em
    disjoint\/} intervals.
\end{description}

We then  have a unique interval associated with each regular path,
and hence with each topological edge of size $> 0$. Moreover, we
choose the minimum of such an interval as a unique number
associated with a topological edge. Specifically, the minimal
number serves as a {\em
  key\/} in a \textit{dictionary}.\index{dictionary} Recall that in general, a dictionary consists
of pairs $\langle\mbox{key},\mbox{item}\rangle$, where the item is
unique for each key. Given a number $k$, the function which
returns the item with the maximal key smaller than $k$ can be
implemented in $O(\log N)$ time, where $N$ is the number of items
in the dictionary \cite{CLR}.

The items we use contain the following information.
\begin{enumerate}
\item An identifier of the topological edge associated with the key.
\item The number of regular vertices on the regular path corresponding
  to this topological edge.
\item An identifier of the regular vertex that has the key as number on this
path.
\end{enumerate}
In Figure~\ref{dict} we give an example of a dictionary containing
three keys, corresponding to the three topological edges in the
simplification $G_s$ of the graph $G$.

\begin{figure}
  \centering
\epsfig{file=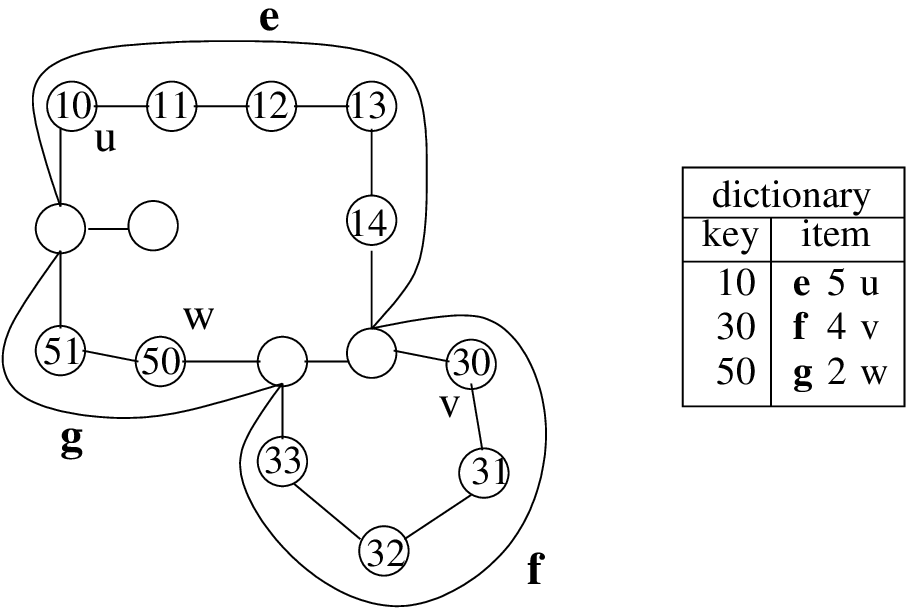} \caption{Dictionary
example.}\label{dict}
\end{figure}

\subsection{Maintaining the numbers of the regular vertices}
\label{secnumbers}

We must now show how to maintain this numbering under
updates, such that the interval and disjointness properties
mentioned above remain satisfied.

Actually, only in case~3 in Section~\ref{sec:update} we need to do
some maintenance work on the numbering. Indeed, by merging two
topological edges, the numbering of the regular vertices is no
longer necessarily consecutive. We resolve this by {\em
renumbering\/} the vertices on the shorter of the two regular
paths. Note that the size of a regular path is stored in the
dictionary item for that path.

In order to keep the intervals disjoint, we must assume that the
maximal number of edge insertions to which we need to respond is
known in advance.  Concretely, let us assume that we have to react
to at most $\ell$ update operations. This assumption is
rather harmless. Indeed, 
  one can set this maximum limit to a large number. If it is eventually
  reached, we restart from scratch.
 A
regular path is ``born'' with at most two regular vertices on it.
Every time a new regular path is created, say the $k$th time, we
assign the number $2k\ell$ to one of the two regular vertices on
it.
 Hence, newly created topological edges correspond to
numbers which are $2\ell$ apart from each other. Since a newly
created topological edge can become at most $\ell-1$ vertices
longer, no interference is possible.

\subsection{Finding the topological edge} Consider that we are in
one of the cases 4--6 described in Section~\ref{sec:update}, where
we have to split the topological edge at
 vertex $x$. We look at the number of $x$, say $k$, and find in
the dictionary the item associated with the maximal key smaller
than $k$. This key corresponds to the interval to which $k$
belongs, or equivalently, to the regular path to which $x$
belongs. In this way we find the topological edge which has to be
split, since this edge is identified in the returned item.

The numbering thus enables us to find an edge in $O(\log
m^\prime)$ time, where $m^\prime$ is the number of edges in $G_s$
which correspond to a regular path passing through at least one
regular vertex. Because $m^\prime$ is at most $m$, the number of
edges in $G$, we obtain:
 \begin{proposition}
   Given a regular vertex and its number, the dictionary returns in
   $O(\log m)$ time the topological edge corresponding to the regular
   path on which this regular vertex lies.
 \end{proposition}

We next show how, when a topological edge is split, we can quickly
find the weights of the two new edges created by the split.

\subsection{Assigning weights to the regular vertices}
The \textit{weight} of a regular vertex $v$ will be denoted by
$\lambda^\ast(v)$.  Weights will be assigned to the regular
vertices such that if $v$ and $w$ are two consecutive regular
vertices with weights $\lambda^\ast(v)$ and $\lambda^\ast(w)$
respectively, then
$\lambda(\{v,w\})=|\lambda^\ast(v)-\lambda^\ast(w)|$.

% We define the \textit{weight} of a regular
%vertex $v$, denoted by $\lambda^*(v)$, as follows.\index{regular
%vertex!weight} Define the {\em $k$th vertex\/} of a regular path
%as the vertex with number $s$, such that $s-s_{\rm min}=k$, where
%$s_{\rm min}$ is the minimal number of the vertices on the regular
%path. Let $v_k$ be the $k$th regular vertex, and let $\{x,y\}$ be
%the topological edge corresponding to the regular path. We now
%define
% $\lambda^\ast(v_0):=0$, and
%$\lambda^\ast(v_k):=\lambda^\ast(v_{k-1})+\lambda(\{v_{k-1},v_k\})$
%for $k>0$.

\subsection{Maintaining the weights of regular vertices}
The maintenance of the weights of regular vertices under edge
insertions is easy. It requires only constant time
 when a topological edge is extended.
Indeed, let $\{x,y\}$ be a topological edge, and suppose that we
extend this edge by inserting $\{y,z\}$. Let $u$ be the regular
vertex adjacent to $y$. Then,
\begin{itemize}
\item if  $\lambda^\ast(u)<0$,
 then $\lambda^\ast(y):=\lambda^\ast(u)-\lambda(\{u,y\})$.
\item if $\lambda^\ast(u)\geqslant 0$, and no regular vertex
with a positive weight is adjacent to $u$, then
$\lambda^\ast(y):=\lambda^\ast(u)+\lambda(\{u,y\})$.
 Otherwise, let $v$ be the regular vertex adjacent to $u$. If
$\lambda^\ast(v) > \lambda^\ast(u)$, then let
 $\lambda^\ast(y) = \lambda^\ast(u) - \lambda(\{u,y\})$, else let
$ \lambda^\ast(y) = \lambda^\ast(u) + \lambda(\{u,y\})$.
\end{itemize}

When a topological edge is split, no adjustments to the weight of the
remaining regular vertices is needed at all. However,
when two topological edges are merged we need to adjust
the weights of the regular vertices on the shortest of the two
regular paths, as shown in Figure~\ref{weights}. This adjustment
of the weights can clearly be done simultaneously with the
renumbering of the vertices.
\begin{figure}
  \centering \epsfig{width=0.8\textwidth,file=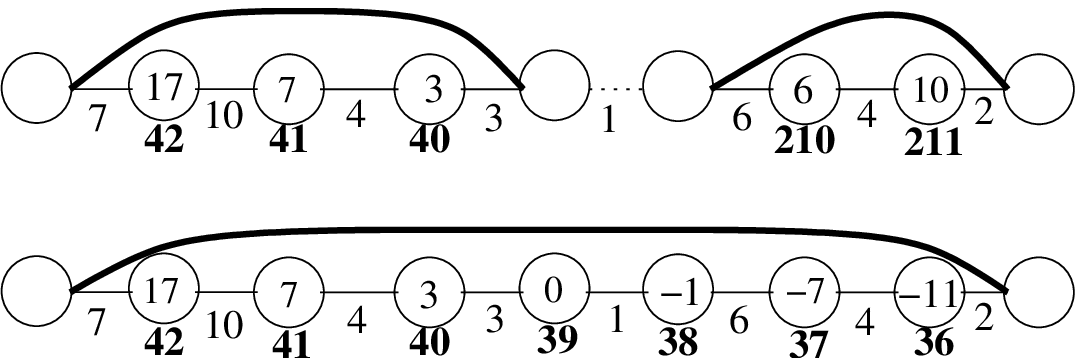}
\caption{Assigning new numbers and weights of regular vertices
  simultaneously when two topological edges are merged. The numbers of
  regular vertices are in bold, the weights are inside the
  vertices.}\label{weights}
\end{figure}

\subsection{Finding the weights} The weights of regular vertices
now enable us to find the weights of the two edges created by a
split of a topological edge in logarithmic time. Indeed, given the
number of the regular vertex where the split occurs, we search in
the dictionary which topological edge needs to be split; call it
$\{z_1,z_2\}$. In the returned item we find the vertex which has
the minimal number of the vertices on the regular path
corresponding to $\{z_1,z_2\}$. Denote this vertex with $u$ which
is adjacent to either $z_1$ or $z_2$. We assume that $u$ is
adjacent to $z_1$, the other case being analogous. The weight of
the two new topological edges $\{z_1,x\}$ and $\{x,z_2\}$ can be
computed easily:
\begin{itemize}
\item
$\lambda(\{z_1,x\}):=\lambda(\{z_1,u\})+|\lambda^\ast(u)-\lambda^\ast(x)|$;
and
\item
$\lambda(\{x,z_2\}):=\lambda(\{z_1,z_2\})-\lambda(\{z_1,x\})$.
\end{itemize}

If only one regular vertex remains on a regular path after a
split, or a regular vertex becomes singular, then the weight of
this vertex is set to $0$. This can all can be done in constant
time, after the topological edge which needs to be split has been
looked up in the dictionary.

\subsection{Complexity analysis} \label{Sec:amortize} By the {\em
amortized complexity\/}\index{amortized complexity} of an on-line
algorithm \cite{Tar,Mehl}, we mean the total computational
complexity of supporting $\ell$ updates (starting from the empty
graph), as a function of $\ell$, divided by $\ell$ to get the
average time spent on supporting one single update. We will prove
here that the Renumbering Algorithm has  $O(\log \ell)$ amortized
time complexity.
 We only count edge insertions because the insertion of
 an isolated vertex has zero cost.

\begin{theorem}\label{thm:renum}
The total time spent on $\ell$ updates by the
  Renumbering Algorithm  is $O(\ell\log \ell)$.
\end{theorem}
\begin{proof} If we look at the general description of the
Renumbering Algorithm, we see that in each case only a constant
number of steps are performed. These are either  elementary
operations on the graph, or dictionary lookups. There is however
one important exception to this. In cases where we need to merge
two topological edges, the renumbering of regular vertices (and
simultaneous adjustment of their weights) is needed. Since every
elementary operation on the graph takes constant time, and every
dictionary lookup takes  $O(\log \ell)$ time, all we have to prove
is that the total number of vertex renumberings is $O(\ell \log \ell)$.

A key concept in our proof is the notion of a {\em super edge}
(see Figure~\ref{fig:super}). Super edges are sets of topological
edges which can be defined inductively: initially each topological
edge (with one or two regular vertices on it) is a member of a
separate super edge. If a member ${\bf a}$ of a super edge ${\bf
A}$ is merged with a member ${\bf
  b}$ of another  super edge ${\bf B}$, then the two super edges are
unioned together in a new super edge ${\bf C}$ and ${\bf a}$ and
${\bf
  b}$ are merged into a new member ${\bf c}$ of the new super edge ${\bf
  C}$. If a member ${\bf d}$ of a super edge is split into ${\bf e}$ and
${\bf f}$, then both ${\bf e}$ and ${\bf f}$ will belong to the
same super edge as ${\bf d}$ did. The important property of super
edges is that the total number of vertices can only grow. We call
this number the {\em size\/} of a super edge.  A split operation
does not affect the size of super edges, while merge operations
can only increase it.

\begin{figure}
\centering \epsfig{width=\textwidth,file=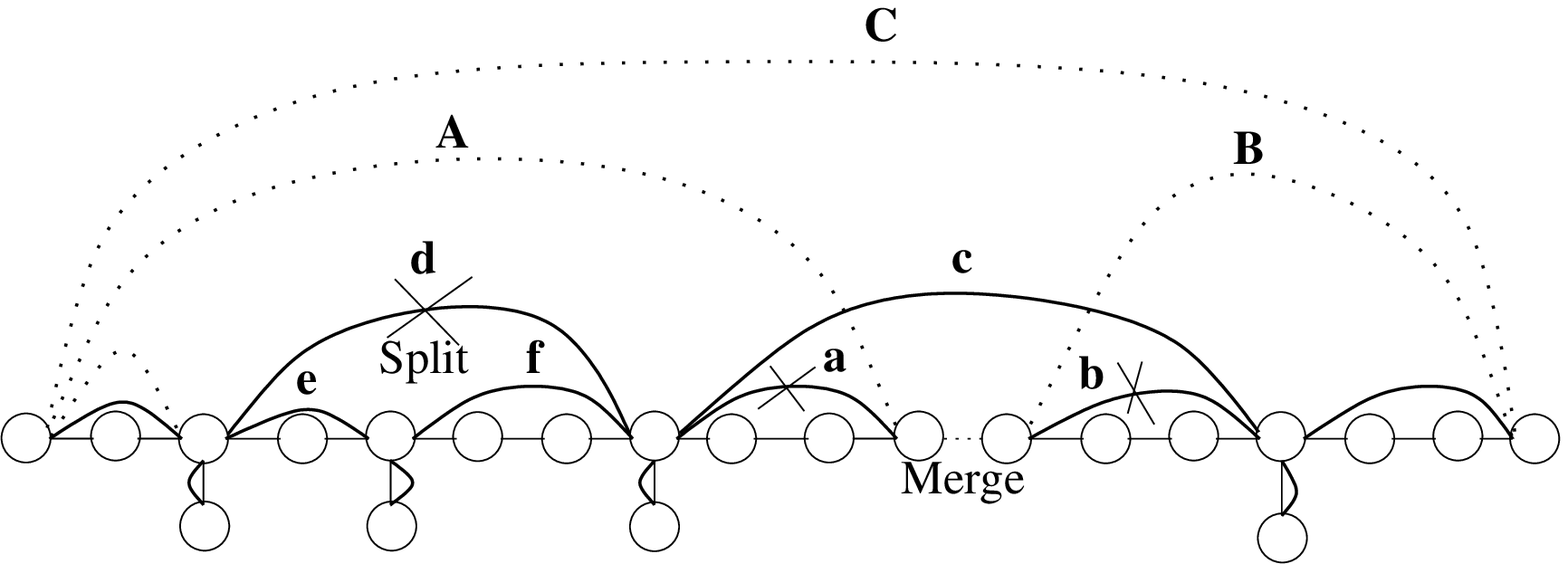} \caption{An
example of some super edges (dotted lines)}\label{fig:super}
\end{figure}

It now suffices to show that
the total number of vertex
renumberings in a super edge of size $\ell$ is $\ell \log \ell$.
We will do this by induction.

The statement is trivial for $\ell=0$, so we take $\ell>0$. We may
assume that the $\ell$th update involves a merge of two
topological edges, since this is the only update for which we have
to do renumbering. Suppose that the sizes of the two super edges
being unioned are $\ell_1$ and $\ell_2$. Without loss of
generality assume that $\ell_1 \leq \ell_2$. Hence,  according to
the Renumbering Algorithm which renumbers the shortest of the two,
we have to do $2\ell_1$ renumbering steps: $\ell_1$ to assign new
numbers, and $\ell_1$ to assign new weights. The size of the new
super edge will be $\ell = \ell_1 +\ell_2$. By the induction
hypothesis, the total number of renumberings already done while
building the two given super edges are $\ell_1 \log \ell_1$ and
$\ell_2 \log \ell_2$. It is known (\cite{Tamassia}) that
\begin{equation}\label{eq:logeq}
2\min\{x,1-x\}\leqslant\ x\log\frac{1}{x}+(1-x)\log\frac{1}{1-x},
\end{equation}
for $x\in[0,1]$. Define $x=\ell_1/\ell$. By~(\ref{eq:logeq}), we
then obtain the inequality
\begin{eqnarray*}
\ell_1 \log \ell_1 + \ell_2 \log \ell_2 + 2\ell_1 & \leqslant &
\ell \log \ell,
\end{eqnarray*}
as had to be shown.
\end{proof}

To conclude this section, we recall from Section~\ref{secnumbers} that
the maximal number assigned to a regular vertex is $2\ell^2$. So,
all numbers involved in the Renumbering Algorithm  take only
$O(\log \ell)$ bits in memory. Theorem~\ref{thm:renum}
assumes the standard RAM  computation model with unit costs. If
logarithmic costs are desired, the total time is $O(\ell \log^2\ell)$.

\section{Online-Simplification: Topology Tree Algorithm}\label{sec:fred} In this
section we introduce another algorithm for keeping the
simplification of a graph up-to-date when this graph is subject to
edge insertions. We only describe the case of edge insertion, but
it is straightforward to extend the Topology Tree Algorithm to a
fully dynamic algorithm, which can also react to deletions.
The algorithm uses a direct adaptation of the topology-tree data structure
introduced by Frederickson~\cite{fred1,fred2}.
This data structure has been used extensively in other
partially and fully dynamic algorithms~\cite{italiano}.

We first show how the topological edge can be found efficiently.

\subsection{Regular multilevel partition} We define a {\em
cluster\/}\index{cluster} as a set of vertices. The {\em size\/}
\index{cluster!size} of a cluster is the number of vertices it
contains. A {\em regular cluster\/}
\index{cluster!regular}\index{regular!cluster} is a cluster of
size at most two,
 containing adjacent regular vertices.
A {\em regular partition\/} of a graph $G$ is a
\index{regular!partition} partition of the set $V_r$ of regular
vertices, such that
 for any two adjacent regular vertices $v$ and $w$, the following holds:
\begin{itemize}
\item either $v$ and $w$ are in the same regular cluster $\mathcal{C}$; or
\item $v$ and $w$ are in different regular clusters $\mathcal{C}_v$ and $\mathcal{C}_w$,
 and at least one of these regular clusters has size two.
\end{itemize}
A {\em regular multilevel partition\/} \index{regular!multilevel
partition}
 of a graph $G$ is a set of partitions of $V_r$ that satisfy the following (see Figure~\ref{fig:regularpart}):
\begin{enumerate}
\item For each level $i=0,1,\ldots,k$, the clusters at level $i$ form a
  partition of $V_r$.
\item The clusters at level $0$ form a regular partition of $V_r$.
\item The clusters at level $i$ form a regular partition when viewing
  each cluster at level $i-1$ as a regular vertex.
\end{enumerate}

\begin{figure}
\centering \epsfig{width=\textwidth,file=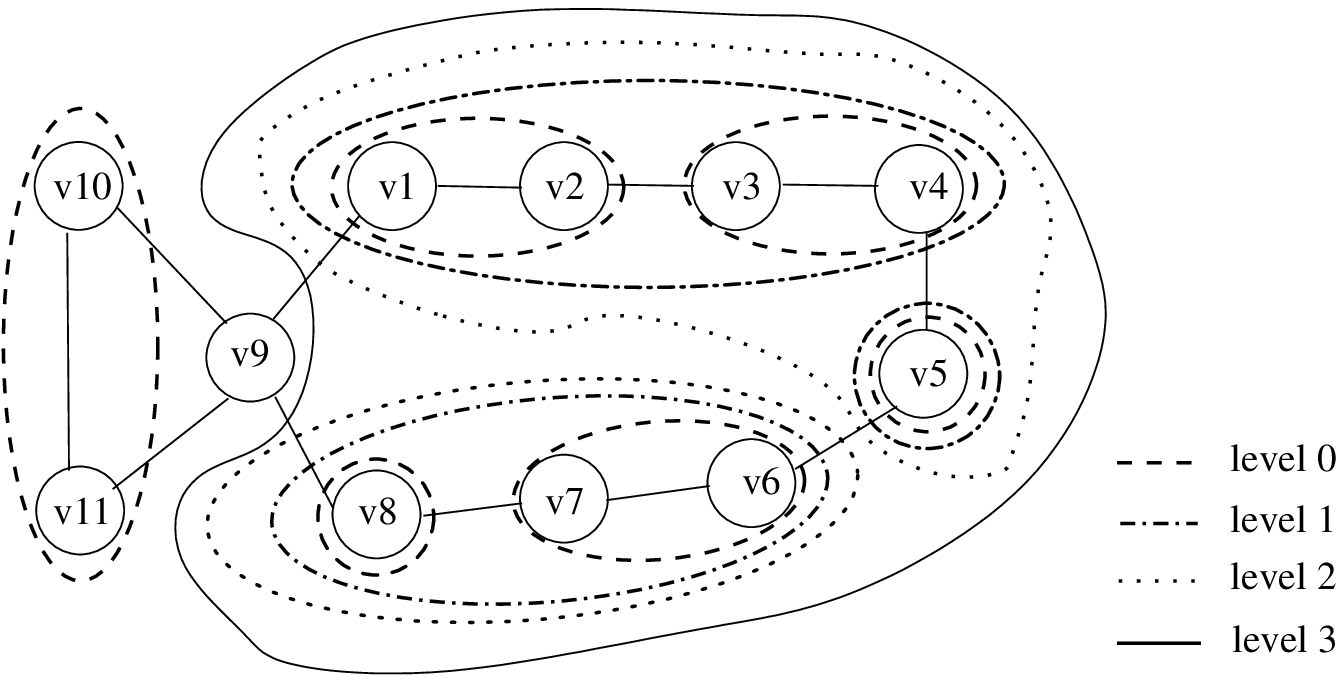}
\caption{Example of a regular multilevel partition of a graph.}
\label{fig:regularpart}
\end{figure}
\begin{figure}
\centering \epsfig{width=\textwidth,file=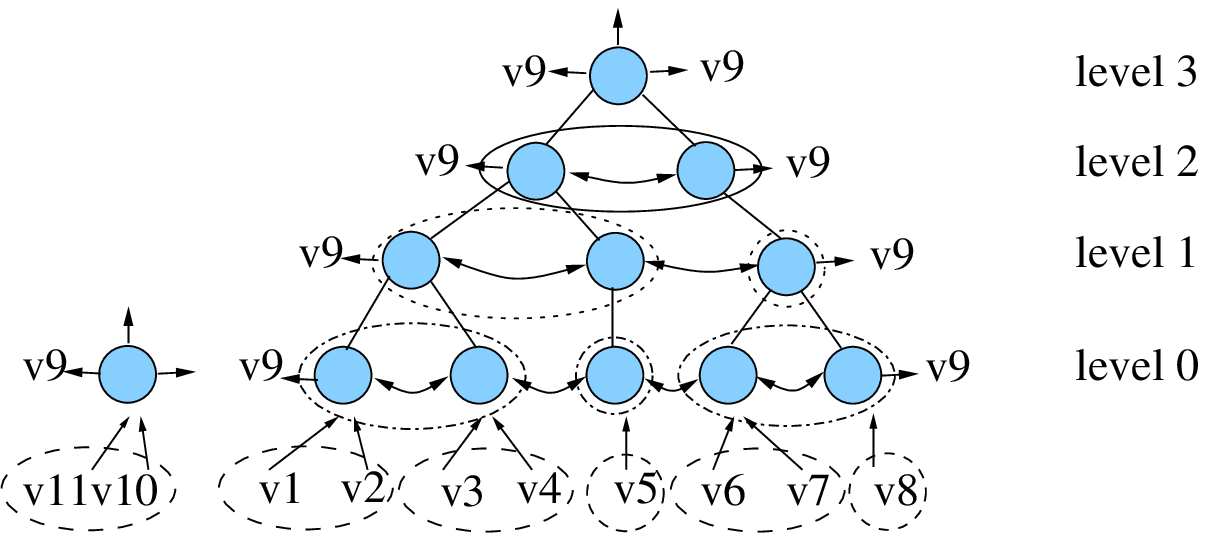}
\caption{The regular forest corresponding to the regular
multilevel partition shown in Figure~\ref{fig:regularpart}.}
\label{fig:regulartree}
\end{figure}

A {\em regular forest\/} \index{regular!forest} of a graph $G$ is
a forest based on a regular multilevel partition of $G$.
We focus on the construction of a single tree in the forest
corresponding to a single regular path. A single tree is
constructed as follows (see Figure~\ref{fig:regulartree}).
\begin{enumerate}
\item A vertex at level $i$ in the tree represents a cluster at level
  $i$ in the regular multilevel partition.
\item A vertex at level $i>0$ has children that represent the clusters
  at level $i-1$ whose union is the cluster it represents.
\end{enumerate}
The height of a topology tree is logarithmic in the number of
regular vertices in the leafs~\cite{fred1}.

We also store adjacency information for the clusters. Two regular
clusters $\mathcal{C}$ and $\mathcal{C}'$ at level $0$ are
\textit{adjacent}\index{cluster!adjacent},
 if there exists a vertex $v\in\mathcal{C}$ and a vertex
$w\in\mathcal{C}'$ such that $v$ and $w$ are adjacent in $G$.

We call two clusters $\mathcal{C}$ and $\mathcal{C}'$ at level $i$
adjacent, if they have adjacent children. A regular cluster
$\mathcal{C}$ at level $0$ is adjacent to a singular vertex $s$ if
there exists a regular vertex $v\in\mathcal{C}$ adjacent to $s$. A
cluster at level $i>0$ is adjacent to a singular vertex $s$ if it
has a child adjacent to $s$.

\subsection{Maintaining a regular multilevel partition}
The following procedure, for maintaining a regular multilevel partition under
edge insertions, closely follows the procedure described by Frederickson
\cite{fred1}, as our data structure is a direct adaptation of Frederickson's.
\paragraph{level $0$}
It is very
easy to adjust the regular partition, i.e., the regular clusters
at level $0$ of the regular multilevel partition. When an edge
$e=\{x,y\}$ is inserted, we distinguish  between the following
cases: 1. the edge $e$ destroys a regular  vertex $u$; 2. the edge
$e$  destroys two regular vertices $u$ and  $v$; 3. the edge $e$
creates a regular vertex  $u$; 4. the  edges $e$ creates two
regular vertices $u$ and  $v$; 5. the  edge $e$ does not change
the number of regular vertices.  We denote with $C_u$ ($C_v$) the
regular cluster containing the vertex $u$ ($v$). We treat these
cases as follows.
\begin{enumerate}
\item  If  the  size  of  $C_u$  is $1$,  then  this  cluster  is
deleted. Otherwise if $C_u$ is adjacent  to a cluster $C$ of size
one, remove $u$ from $C_u$ and union $C_u$ with $C$.
\item Apply case 1 to both $C_u$ and $C_v$.
\item Create a new cluster $C_u$ only containing $u$. If $C_u$ is
adjacent to a cluster $C$ of size one, union $C_u$ with $C$.
\item Apply case 3,  but if  both $C_u$  and $C_v$  are not
adjacent to a  cluster of size one, then they are unioned
together.
\item Nothing has to be done.
\end{enumerate}

As an example consider the graph depicted in
Figure~\ref{fig:regcluster}.  The insertion of edge $\{x,y\}$
destroys the regular vertex $x$, so we are in case 1. Because
$\mathcal{C}'$ is adjacent to $\mathcal{C}''$ and the size of
$\mathcal{C}''$ is one, we must union $\mathcal{C}'$ and
$\mathcal{C}''$ together into a new regular cluster $\mathcal{C}$.
The maintenance of the regular partition is completed after
adjusting the adjacency information of both $\mathcal{C}$ and
$\mathcal{D}$, as shown in Figure~\ref{fig:regcluster}.
\begin{figure}
\centering \epsfig{height=5cm,file=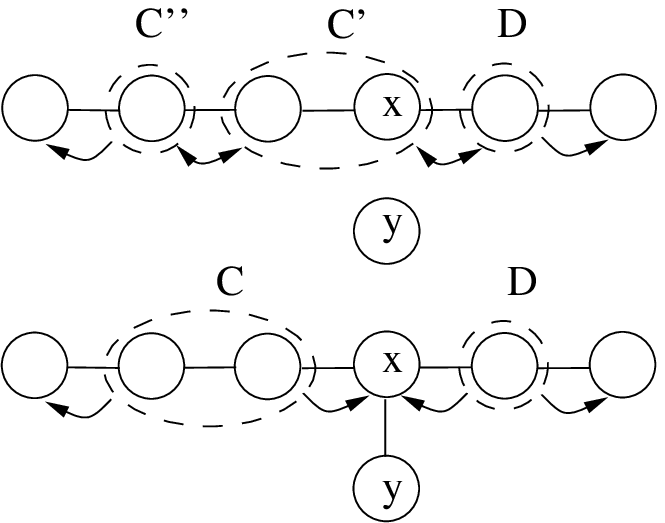}
\caption{Adjusting the regular partition after inserting edge
  $\{x,y\}$.}\label{fig:regcluster}
\end{figure}

\paragraph{level $> 0$}
We assume that the regular partition at level $0$ reflects the
insertion of an edge, as discussed above. The number of clusters
which have changed, inserted or deleted is at most some constant.
We put these clusters in a list $L_C$, $L_I$, and $L_D$ according
to whether they are changed, inserted or deleted. More
specifically, these lists are initialized as follows. Each regular
cluster that has been split or combined to form a new regular
 cluster is inserted in $L_D$, while each new regular cluster
is inserted in list $L_I$. The adjacency information is stored
with the clusters in $L_I$. For clusters in $L_D$ every adjacency
information is set to null, except the parent information. For
each regular cluster whose set of vertices has not changed but its
adjacency information has changed, update the adjacency
information and insert it into $L_C$.

We create lists $L_D'$, $L_I'$, and $L_C'$ to hold the clusters at
the next higher level of the regular multilevel partition. These
lists are initially empty.

We first adjust the clusters in the list $L_D$. Every cluster
$\mathcal{C}$ in $L_D$ is removed from $L_D$, and $\mathcal{C}$ is
removed as child from its parent $\mathcal{P}$ (if existing).
\begin{itemize}
\item If $\mathcal{P}$ has no more children, then insert $\mathcal{P}$ in
  $L_D'$.
\item If $\mathcal{P}$ still has a child $\mathcal{C}'$, then if
  $\mathcal{C}'$ is not already in $L_C$ or $L_D$, then insert
  $\mathcal{C}'$ into $L_C$.
\end{itemize}
Next, we search the list $L_C$ for clusters that have siblings.
Suppose that $\mathcal{C}\in L_C$ has a sibling $\mathcal{C}'$ and
parent $\mathcal{P}$.
\begin{itemize}
\item If $\mathcal{C}$ and $\mathcal{C}'$ are adjacent, then remove
  $\mathcal{C}$ from the list $L_C$, and  remove $\mathcal{C}'$ from
  $L_C$ if it is in this list. Insert $\mathcal{P}$ into $L_C'$.
\item If $\mathcal{C}$ and $\mathcal{C}'$ are not adjacent, then remove
  $\mathcal{C}$ and $\mathcal{C}'$ as children from $\mathcal{P}$.
  Remove $\mathcal{C}$ from the list $L_C$, and also remove
  $\mathcal{C}'$ from $L_C$ if it is in this list. Insert both
  $\mathcal{C}$ and $\mathcal{C}'$ into $L_I$, and insert $\mathcal{P}$
  in $L_D'$.
\end{itemize}

Finally, we treat the remaining clusters in $L_C$ and in $L_I$.
Let $\mathcal{C}$ be such a cluster. Remove $\mathcal{C}$ from the
appropriate list. In what follows, the degree of $\mathcal{C}$ is
the number of adjacent clusters.
\begin{itemize}
\item If $\mathcal{C}$ has degree zero, then it is the root of a tree in
  the regular forest. Insert its parent $\mathcal{P}$, if existing,
  in $L_D'$.
\item If $\mathcal{C}$ has degree one or two, then we have the following
  possibilities:
\begin{itemize}
\item If every adjacent cluster to $\mathcal{C}$ has a sibling, then
  insert the parent $\mathcal{P}$ of $\mathcal{C}$ into $L_C'$ in case
  $\mathcal{P}$ exists. In case  $\mathcal{C}$ does not have a parent,
  create a new parent cluster $\mathcal{P}$ and insert it into
  $L_I'$.
\item Let $\mathcal{C}'$ be a cluster adjacent to $\mathcal{C}$ which
  has no sibling. Remove $\mathcal{C}'$ from the appropriate list,
  if it is in  a list. If both $\mathcal{C}$ and $\mathcal{C}'$ have a parent,
  denoted by $\mathcal{P}$ and $\mathcal{P}'$ respectively, then remove
  $\mathcal{C}$ as child of $\mathcal{P}$ and make it a child of
  $\mathcal{P}'$. Insert $\mathcal{P}$ into $L_D'$, and insert
  $\mathcal{P'}$ into $L_C'$. If both $\mathcal{C}$ and $\mathcal{C}'$
  have no parent, then create a new parent $\mathcal{P}$ of
  $\mathcal{C}$ and $\mathcal{C}'$, and insert $\mathcal{P}$ into $L_I'$.
  If $\mathcal{C}$ has a parent $\mathcal{P}$, and $\mathcal{C}'$ has no
  parent, then make $\mathcal{C}'$ a child of $\mathcal{P}$ and insert
  $\mathcal{P}$ into $L_C'$. The case that $\mathcal{C}'$ has a parent
  $\mathcal{P}'$, and $\mathcal{C}$ has no parent, is analogous.
\end{itemize}
\end{itemize}
When all clusters are removed from $L_D$, $L_C$, and $L_I$,
determine and adjust the adjacency information for all clusters in
$L_D'$, $L_C'$, and $L_I'$ and reset $L_C$ to be $L_C'$, $L_C$ to
be $L_C'$, and $L_I$ to be $L_I'$. If no clusters are present in
$L_D'$, $L_C'$ or $L_I'$, nothing needs to be done and the
iteration stops. This completes the description of how to handle
the lists $L_D$, $L_C$, and $L_I$.

\subsection{Finding a topological edge} Consider that we are in one
of the cases 4--6 described in Section~\ref{sec:update}, where we
have to split a topological edge. Let $x$ be the regular vertex at
which we have to split the topological edge. We store a pointer
from $x$ to the regular cluster $\mathcal{C}_x$ in which it is
contained. We also store a pointer from each root of a tree $T$ in
the regular forest to the topological edge, corresponding to the
regular path formed by all vertices in the leaves of $T$. We find
the topological edge which needs to be split by going from
$\mathcal{C}_x$ to the root of the tree containing
$\mathcal{C}_x$. Since the height of the tree is at most $O(\log
\ell)$, where $\ell$ is the current number of edge insertions, we
obtain the following.
 \begin{proposition}
   Given a regular vertex $x$, the regular forest returns  the topological edge corresponding to the regular path on
   which this regular vertex lies in $O(\log \ell)$ time.
 \end{proposition}

\subsection{Storing weight information} We store weight
information in two different places. We define the weight of a
regular cluster at level $0$ of size one as zero. Let
$\mathcal{C}$ be a cluster at level $0$ of size two, and let $v$
and $w$ be the two regular vertices in $\mathcal{C}$. Then we
define the weight of $\mathcal{C}$ as the weight of the edge
$\{v,w\}$. If a cluster at level $0$ is adjacent to a singular
vertex $s$, then we store the weight of $\{v,s\}$ together with
the adjacency information (here, $v$ is the vertex in
$\mathcal{C}$ adjacent to $s$). If two clusters $\mathcal{C}$ and
$\mathcal{C}'$ at level $0$ are adjacent, then we store the weight
of $\{v,w\}$ together with their adjacency information (here
$v\in\mathcal{C}$ and $w\in\mathcal{C}'$ and $v$ is adjacent to
$w$).

The weight of a cluster of size one at level $i>0$,
 is defined as the weight of its child at the next lower level.
  The weight of a cluster of size two  at level $i>0$ equals
 the sum of the weights of its two children and the weight
   stored with their adjacency information. If two
clusters at level $i>0$ are adjacent, we store the weight of the
adjacency information of their adjacent children. If a cluster at
level $i>0$ is adjacent to a singular node, we store the weight of
the adjacency information of its child and the singular node.

\subsection{Maintaining weight information} The weight of clusters
and the weights stored together with the adjacency information, is
updated after each run of the update procedure for the regular
multilevel partition,
 with an  extra constant cost.
Indeed, both the weights of clusters at level $0$ and the weights
stored with the adjacency information, are trivially updated. When
we assume that  all levels lower than $i$ represent the weight
information correctly, the weight information of clusters in $L_C$
and $L_I$ is trivially updated using the weight information at
level $i-1$.

\subsection{Finding the weights} As mentioned above, each root of
a regular tree in the regular forest, has  a pointer to a unique
topological edge. This root has its own weight, as defined above,
and is adjacent to two singular vertices. The weight of the
topological edge is obtained by summing the weight of the root
together with the weights of the adjacency information of  the two
singular vertices. This is illustrated in
Figure~\ref{fig:examplewithweights}.
\begin{figure}
  \centering
\epsfig{width=\textwidth,file=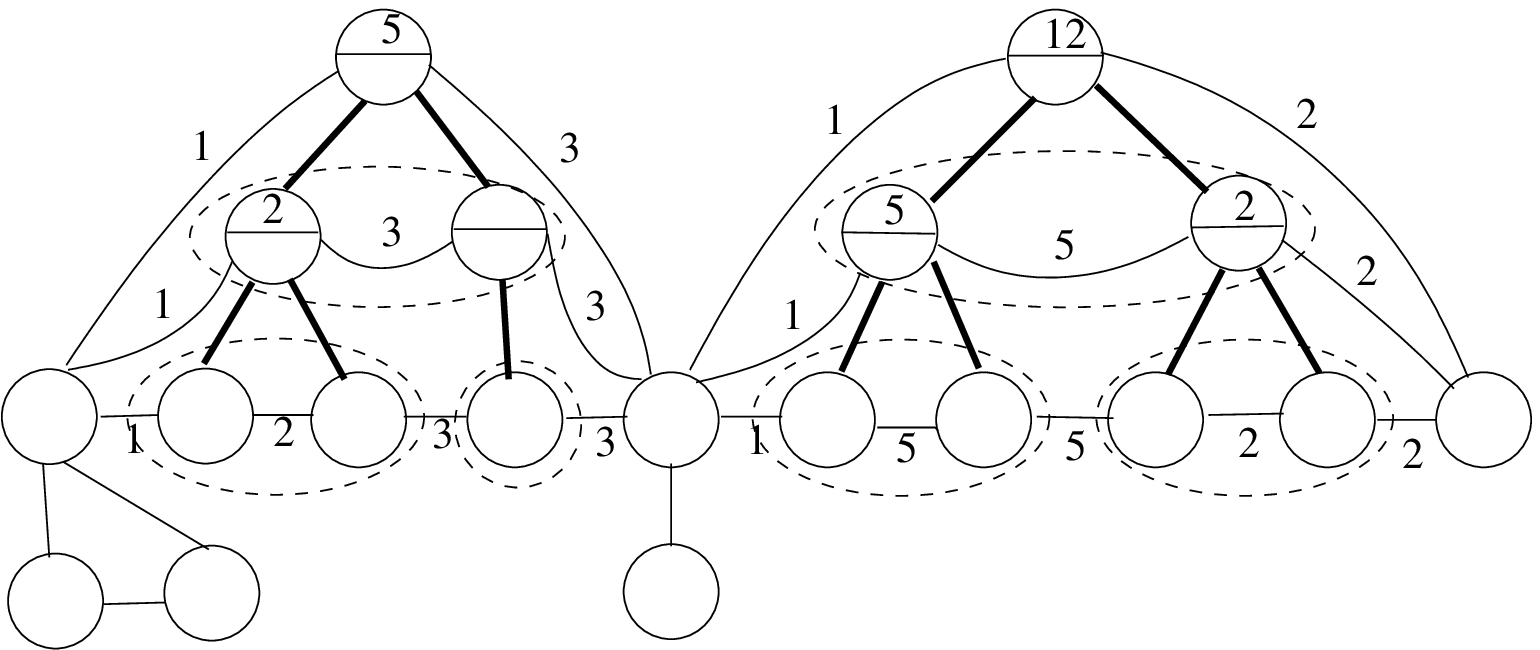}
\caption{Example of a regular tree together with its weight
information.} \label{fig:examplewithweights}
\end{figure}
\subsection{Complexity Analysis} The complexity of the Topology
Tree Algorithm is governed by two things: the maximal height of a
single tree in the regular forest, and the amount of work that
needs to be done at each level in the maintenance of the regular
multilevel partition. We already saw that
the height of a single tree is
logarithmic in the number of regular vertices on the regular
path on which the tree is built. Moreover,
Frederickson has proven that in the lists $L_C$, $L_D$, and
$L_I$ only a constant number of clusters are stored \cite{fred1}.
These lists are updated at most $O(\log \ell)$ times,
where $\ell$ is the number of edge insertions, so that the total update
time is $O(\log \ell)$ per edge insertion. Hence, we may
conclude the following:
\begin{theorem} The total time spent on $\ell$ updates by the
Topology Tree Algorithm is $O(\ell\log \ell)$.
\end{theorem}

\section{Experimental Comparison}\label{sec:exp}

The Renumbering Algorithm and the Topology Tree Algorithm are very different,
but have the same theoretical complexity. Hence, the question arises how they
compare experimentally. In this section we try to obtain some insight into
this question.

Both algorithms were implemented in  C++ using LEDA \cite{LEDA}.
We used the GNU g++ compiler version 2.95.2 without any
optimization option. Our experiments were performed on a SUN Ultra
10 running at  440~Mhz with 512~MB internal memory. Implementing
the Renumbering algorithm was considerably easier than
implementing the Topology Tree Algorithm.

We conducted our experiments on
three types of inputs. First of all, we extensively studied
 random inputs, which are random sequences of updates
on random graphs.
Next, we used two kinds of non-random graph inputs
which focus specifically on the merging and the splitting of topological
edges.  Thereto,
we constructed an input sequence which repeatedly
merges topological edges, and an input sequence which first creates a
very large number
 of small topological edges, and then splits these edges randomly.
Finally, we ran both algorithms on two inputs
originating from real data sets.

\paragraph{Methodology} Since the experiments have an element of randomness,
we show the results in the form of 95\% confidence intervals. For each test,
we perform a large number of runs. For each run, we compute the ratio
between the total time taken
by Topology Tree and that taken by Renumbering.
We took the average of these ratios and computed the
$95\%$ confidence interval. So, for example, the interval $[1.10,1.15]$ means
that Topology Tree was 10 to 15\% slower than Renumbering in 95\% of the runs
in the test.

\paragraph{Random Inputs}
The random inputs consist of random graphs that are generated,
given the number of vertices and edges. Each run builds a
random graph incrementally with the  insertions uniformly
distributed over the set of edges. We conducted a series of tests
for different number of nodes $n$ and number of edges $m$. For
every pair of values for $n$ and $m$ we did $1000$ runs. The
results of these experiments are shown in Table~\ref{tabel:een}.
\begin{table}
\begin{center}
\begin{tabular}{|c|c|c|c|}\hline
vertices$\backslash$edges & m=5\,000 & m=10\,000   & m=20\,000
\\\hline n=1\,000 & $[1.10,1.15]$& $[1.03,1.06]$
&$[0.97,0.99]$\\\hline\hline
 vertices$\backslash$edges & m=5\,000 & m=25\,000 &
m=75\,000 \\\hline n=5000 & $[1.25,1.29]$ & $[1.01,1.03]$ &
$[0.96,0.98]$
\\\hline\hline
 vertices$\backslash$edges & m=10\,000 & m=50\,000& m=150\,000 \\\hline
n=50\,000 & $[1.30,1.35]$ & $[1.06,1.07]$ & $[0.91,0.92]$
\\\hline\hline
 vertices$\backslash$edges & m=10\,000 & m=100\,000 & m=300\,000\\\hline
 n=100\,000 & $[1.21,1.23]$ & $[0.98,0.99]$& $[0.85,0.86]$ \\\hline
 \end{tabular}
 \end{center}
 \caption{95\% confidence intervals on ratio between Topology Tree and
 Renumbering, from 1000 runs on random inputs.}
\label{tabel:een}
\end{table}

For small numbers of edge insertions, i.e., when the probability
of having many regular vertices is large, we see that the
Renumbering Algorithm is faster. However, when the number of edge
insertions increases, the Topology Tree Algorithm becomes slightly
faster. This is probably due to the fact that the dictionary in
the Renumbering Algorithm becomes very large, i.e., there are many
short topological edges, and hence it takes longer to search for
topological edges.

\paragraph{Non-Random Inputs}
The non-random inputs consisted of two types. For the first type,
we created  a large number of topological edges and then
started to merge these edges pairwise. The end result was a very
long topological edge. For the second type, we first created a
very large number of short regular paths
consisting of a single regular vertex, and then started to
split these randomly. Each
result shown in Table~\ref{tabel:twee} is obtained from
$100$ runs.
\begin{table}
\begin{center}
\begin{tabular}{|l|c|}\hline
Merge ($n=20\,099$, $m=20\,098$) &
$[3.60,3.74]$\\\hline
Split ($n=280\,000$, $m=200\,000$) &
$[1.15,1.17]$ \\\hline
 \end{tabular}
 \end{center}
 \caption{95\% confidence intervals on ratio between Topology Tree and
 Renumbering, from 100 runs on non-random inputs.} \label{tabel:twee}
 \end{table}

The first type of input was designed in order to reproduce the cases,
observed in the random inputs, where Renumbering is much faster than
Topology Tree.  This is confirmed by the experimental result.  Indeed,
on this type of inputs, the Topology Tree Algorithm has to maintain
large topology trees, which is probably the reason that it is slower.

The second type of input was designed in an attempt to reproduce the
cases where Topology Tree is faster than Renumbering.  Our attempt
failed, however, as the experimental result does not confirm this.
Indeed, although the topology trees all have  height one, while the
dictionary is very large, the Renumbering Algorithm nevertheless still
is faster.

\paragraph{Real Data Inputs}
We also tested the relative performance of both algorithms with
respect to graphs representing real data. We present the results
on two data sets:
\begin{description}
\item[{\rm\em Hydrography graph}] A data set representing the
hydrography of Nebraska. This set contains $157\, 972$ vertices,
of which $96\, 636$ are regular.
\item[{\rm\em Railroad graph}] A
data set representing all railway mainlines, railroad yards, and
major sidings in the continental U.S. compiled at a scale $1:100\,
000$. It contains $133\, 752$ vertices of which only $14\, 261$
are regular. It is available at the U.S. Bureau of Transportation
Statistics (www.bts.gov/gis).
\end{description}
The results shown in Table~\ref{tabel:drie} are obtained after
performing  $100$ experiments. In each experiment, we ran both
algorithms in a random way on these data sets. We computed the
ratio between the total time the Topology Tree Algorithm
 needed to perform the test and the total time the Renumbering Algorithm needed
  to accomplish the same task.
We took the average of this ratio and computed the $95\%$
confidence interval.
\begin{table}
\begin{center}
\begin{tabular}{|l|c|}\hline
Hydrography & $[1.62,1.66]$\\\hline
Railroad & $[0.95,0.96]$ \\\hline
 \end{tabular}
 \end{center}
 \caption{95\% confidence intervals on ratio between Topology Tree and
 Renumbering, from 100 runs on real datasets.}
 \label{tabel:drie}
 \end{table}
Again, we see that when there are only few, but long, topological
edges, the Renumbering Algorithm is faster than the Topology Tree
Algorithm. When there are many, short, topological edges, like in
the railroad graph, the Topology Tree Algorithm is slightly faster
than the Renumbering Algorithm.

\bigskip In summary, our experimental study shows that when the
percentage of regular vertices is high in a graph, then the
Renumbering Algorithm is clearly better than the Topological Tree
Algorithm, and when the same percentage is low, then the reverse
often holds. However, our experimental study did not compare any
specific problem solving with and without using topological
simplification. Intuitively, the value of topological
simplification should increase with the percentage of regular
vertices in the graph. Therefore, when the percentage of the
regular vertices is high, the Renumbering Algorithm should be not
only better than the Topological Tree Algorithm but also yield a
significant time saving over problem solving without topological
simplification. We expect this to be the most important practical
implication of our study for the case when there are only
insertions of edges and vertices into the graph.  However, when a
fully dynamic structure is needed, then the Topological Tree
Algorithm should be also advantageous in practice.

\section*{Acknowledgement}
We would like to thank Bill Waltman for providing us with the hydrography data set.

\end{document}